\documentstyle[12pt,worldsci]{article}
\begin{document}
\def\z2{\ifmmode Z_2\else $Z_2$\fi}
\def\ie{{\it i.e.},}
\def\eg{{\it e.g.},}
\def\etal{{\it et. al.}}
\def\to{\rightarrow}
\def\tcv{\ifmmode t\to cV\else $t\to cV$\fi}
\def\Re{{\cal R \mskip-4mu \lower.1ex \hbox{\it e}\,}}
\def\Im{{\cal I \mskip-5mu \lower.1ex \hbox{\it m}\,}}
\pagestyle{empty}
\setlength{\baselineskip}{2.6ex}

\title{{\bf EXTRACTION OF $Z'$ COUPLING DATA FROM $Z' \to jj$ AT THE LHC AND
SSC}}
\author{T.G.~RIZZO\\
\vspace{0.3cm}
{\em High Energy Physics Division, Argonne National Laboratory, Argonne, IL
60439, USA}}
\maketitle

\begin{center}
\parbox{13.0cm}
{\begin{center} ABSTRACT \end{center}
{\small\hspace*{0.3cm}
A recent analysis has shown that it may be possible at the SSC to extract
information about $Z'$ couplings via the decay $Z' \to jj$. This technique
was found to be useful for some extended electroweak models provided the
$Z'$ is relatively light. In the present paper, we generalize this procedure
to the LHC and to $Z'$'s which are more massive than 1 TeV.}}
\end{center}

Probing the nature of a newly discovered particle at a hadron supercollider
can be a difficult problem and one that is critical to address. For example,
if a Higgs-like object is discovered it will be extremely important to
determine if it is the conventional Higgs of the Standard Model(SM), one of
the Higgs' of the Minimal Supersymmetric Standard Model, or some other more
exotic beast. A similar situation would apply to the discovery of a new gauge
boson($Z'$)-to try and identify {\it which} $Z'$ has been discovered. This
issue has attracted much attention in the literature{\cite {1}} during the
past fews years with various techniques being proposed to extract information
on the $Z'$'s couplings to fermions. Since all of these schemes suffer from
some form of weakness it is clearly of some importance to have as much
artillery available as possible when assaulting the $Z'$.

It has recently been shown that it may be possible, at least for relatively
light $Z'$'s arising from certain classes of extended electroweak models(EEM),
to use the $Z' \to jj$ mode as a potential source
of coupling data{\cite {2,3}}.
The main difficulty with this channel is the enormous background which arises
from QCD even after very tight selection cuts are applied to the data in the
dijet invariant mass range which the $Z'$ is
already known to occupy. Sufficient statistical power must be available to fit
the dijet mass distribution quite precisely outside the signal region before a
background subtraction can be performed. Only then is it possible to have any
hope of seeing excess events due to the $Z'$,  provided of course that
the $Z'$'s
couplings are sufficiently strong. The usefulness of the dijet channel to
probe the $Z'$'s couplings can be quantified by the resulting statistical
significance, $S/{\sqrt {B}}$, of the $Z'$ peak. The purpose of the present
work is to extend this previous analysis to both the LHC and to $Z'$'s with
larger masses. We will see that the canonical order of magnitude higher
integrated luminosity available at the LHC will allow the dijet channel to be
a useful probe of $Z'$ couplings for a much larger range of masses than does
the SSC.

We begin by a quick overview of the analysis as presented in Ref.~3. We assume
that a $Z'$ has already been discovered via it's leptonic modes so that it's
mass and width are relatively well determined. We remind the reader that for
most EEM the $Z'$'s width to mass ratio is usually rather small,
$\Gamma/M_{Z'} \leq 0.05$, so that excess dijets from the $Z'$ will occupy a
rather narrow invariant mass range. To reduce QCD backgrounds we demand that
both jets are very central and have high $p_t$'s, \ie
{}~$-1 \leq \eta_{j_1,j_2} \leq 1$ and $p_t \geq 0.2M_{Z'}$ and we concentrate
on the data in the dijet mass range near the $Z'$, \ie
{}~$0.7 \leq x_{jj} \leq 1.5$, where $x_{jj}=M_{jj}/M_{Z'}$ with $M_{jj}$ being
the dijet invariant mass. For smaller values of $x_{jj}$, outside the above
range, the shape of the mass distribution is perturbed significantly by our
cuts while for larger values of $x_{jj}$ there is a loss is statistics.
Since no real data
exists, both signal and background are generated numerically using a improved
Born calculation{\cite {4}} for the QCD dijet background and a two-loop,
QCD-corrected `K-factor' for the $Z'$ production process{\cite {5}}. QCD
corrections to the $Z'$ decay were also included and several different NLO
parton distributions were employed to ascertain the sensitivity of the results
to variations in these distributions. Both the signal and background were
smeared assuming a dijet mass resolution of $\Delta M_{jj}/M_{jj}=0.034$,
integrated over bins of width $0.025M_{Z'}$, following the ATLAS analysis
{\cite {2}}, and provided with Gaussian statistical fluctuations. Since almost
all of the $Z'$-induced dijets should lie within the range
$M_{Z'} \pm 2\Gamma$, we define the range $0.9 \leq x_{jj} \leq 1.1$ to be the
signal regime and fit the `data' outside this range by a degree-7 polynomial
(once it is rescaled by a factor of $x_{jj}^5$). Polynomials of higher degree
fail to improve the $\chi^2/d.o.f.$ of the fit. The fitted background is then
extrapolated into the signal regime and subtracted from the `data' leaving a
potential $Z'$-induced event excess. This excess dijet distribution is then
fit to either a Gaussian or Breit-Wigner shape and integrated to determine the
total number of $Z' \to jj $ events. Clearly, if the number of signal events
is too small in comparison to the background no obvious excess will be
observed. Since the total number of events is sensitive to a
number of overall systematic uncertainties (\eg ~the integrated luminosity and
the choice of parton distributions) as well as being sensitive to what we
assume the $Z'$ can to, we will normalize the number of $Z' \to jj$ events we
find to
the number of $Z'$-induced dilepton events in the discovery channel which
defines the ratio $R$. (These leptons are assumed to have rapidities in the
range $-2.5 \leq \eta \leq 2.5$.) If $S/{\sqrt {B}}$ is too small, $R$
will suffer from large errors and we will learn little or nothing about the
$Z'$'s couplings.

Fig.~1 shows two examples of where this technique works quite well for a 1 TeV
$Z'$ at the SSC assuming an integrated luminosity of 10 $fb^{-1}$, \ie  ~for
the Left-Right Model(LRM){\cite {6}} with $\kappa=g_R/g_L=1$
and a $Z'$ with SM-like couplings(SSM). For the LRM(SSM) the extracted value
of $R$ from the `data' is $34.9 \pm 4.0$($20.4 \pm 2.2$) while theory predicts
30.5(18.9). In the LRM case, this converts to the $95\%$ CL bound on the
parameter $\kappa$:
$0.83 \leq \kappa \leq 1.11$. Of course, the method works well only because
the statistical significance of the $Z'$ dijet peak is quite high,
$S/{\sqrt {B}} >7$, for these two particular cases. For other models one
finds that $S/{\sqrt {B}}$ is much smaller even for much greater integrated
luminosities. This arises mainly from the fact that for most models the $Z'$
couplings to fermion pairs is somewhat smaller than in either the LRM or SSM
examples. The Alternative version of the Left-Right Model(ALRM) and the
$E_6$ effective Rank-5 models(ER5M){\cite {7,8}}, which are
described by a parameter $\theta$, are reasonably representative of models in
this class. For all values of $\theta$ one finds that a 1 TeV $Z'$ at the SSC
would be essentially impossible to observe in the dijet channel unless the
integrated luminosity was significantly larger than 10 $fb^{-1}$. Fig.~2 shows
this explicitly for the case $\theta=-\pi/2$, which is usually referred to as
model $\chi$ in the literature. Among the ER5M, $\chi$ has essentially the
largest dijet cross-section which implies that for other values of $\theta$
the situation can be significantly worse. The $Z'_{\chi}$ peak is not visible
with only 10 $fb^{-1}$ but is much more respectable for 100 $fb^{-1}$ of
luminosity. However, even in this case, the extracted value of $R$ from the
`data', $R=12.7 \pm 2.7$, is found to not only agree with the theoretical
prediction for this model, $R=9.6$, but with the predictions of {\it {all}}
ER5M with $\theta$'s outside the range $9^{\circ} \leq \theta \leq 39^{\circ}$.
Thus although the increased integrated luminosity has helped us to observe the
$Z'$, it's not sufficient to provide us with a precise enough determination
of $R$ which we need for model discrimination. Clearly, this implies that a
value of $S/{\sqrt {B}}>5-6$ is a {\it {minimum}} requirement to use this
technique.

If we use this minimal criterion as a guidepost for our ability to use $R$ as
a model discriminator, we can ask how well our procedure works for other
models, at the LHC, or for more massive $Z'$'s. These possibilities are
addressed by the results shown in Figs.~3a-f and Figs.~4a-b to which we now
turn. From Fig.~3a we see that the dijet analysis can be applied to a 2 TeV
LRM $Z'$ at the SSC provided the integrated luminosity available is increased
to about 25 $fb^{-1}$. $Z'$'s of somewhat greater mass would appear to be
quite hopeless requiring more than 10 standard years of running to accumulate
adequate statistics. At the LHC, however, we see from Fig.~3b that the
factor of 10 larger design luminosity may allow us to use $R$ as a model
discriminator for masses approaching 3 TeV in the LRM case after a few years
of running. (It is important to note that the slopes of the LHC curves are
steeper than those for the SSC due to the LHC's lower value of ${\sqrt {s}}$.)
Figs.~3c-d show a very similar story for the SSM $Z'$ since its production
cross section is comparable to but slightly larger than that for the LRM. For
the ALRM $Z'$ case, shown in Figs.~3e-f, the situation is entirely different
however. We see that $R$ can probably never be determined at the SSC, even for
a $Z'$ mass of 1 TeV, due to the small cross section (although an upper bound
might be obtainable). At the LHC, a 1 TeV $Z'$ arising from this model might
be probed after several years of running but for larger masses our dijet
technique will surely fail.

The situation for the ER5M is not qualitatively different from the ALRM case,
as one might expect, but is still somewhat sensitive to the value of the
parameter $\theta$. For the $\chi$-type $Z'$, we see from
Fig.~4a that the SSC with an integrated luminosity of 100 $fb^{-1}$ just
barely manages to satisfy our `minimal' criteria constraint, which is why $R$
was perhaps not as precisely determined as well as we would have liked in the
discussion above. Larger $Z'$ masses are clearly hopeless at the SSC. At the
LHC, from Fig.~4b, we see that the couplings of a 1 TeV $Z'_{\chi}$ has a
reasonably good chance of being probed by the present dijet analysis after
only 2-3 years of running at the canonical luminosity. Larger masses seem to
be essentially impossible. As noted above, the $\chi$ case is realistically the
most optimistic of all the ER5M. To show this explicitly, we consider a
different ER5M which has often been discussed in the literature, called $\eta$.
(This corresponds to choosing the parameter $\theta=cos^{-1}{\sqrt {5/8}}$.)
Figs.~4c-d show us directly that for a 1 TeV $Z'_{\eta}$, neither collider
will be able provide us with coupling information with less than a decade of
running! This clearly demonstrates the shortfall of this technique, \ie ~it can
only be applied for relatively light $Z'$'s and even then only for certain
classes of EEM in which the $Z'$ has relatively strong couplings to fermion
pairs.

Once a new particle is produced at the SSC/LHC, our work is just beginning. We
must go beyond discovery and be able to determine just what it is that has
been found.
Although the procedure that we've described above cannot be used for a $Z'$
originating from an arbitrary EEM if it is overly massive, it does add an
important ingredient into the mix of techniques with which the $Z'$'s
couplings can be probed at hadron supercolliders.

\vspace{1.0cm}
%
\def\MPL #1 #2 #3 {Mod.~Phys.~Lett.~{\bf#1},\ #2 (#3)}
\def\NPB #1 #2 #3 {Nucl.~Phys.~{\bf#1},\ #2 (#3)}
\def\PLB #1 #2 #3 {Phys.~Lett.~{\bf#1},\ #2 (#3)}
\def\PR #1 #2 #3 {Phys.~Rep.~{\bf#1},\ #2 (#3)}
\def\PRD #1 #2 #3 {Phys.~Rev.~{\bf#1},\ #2 (#3)}
\def\PRL #1 #2 #3 {Phys.~Rev.~Lett.~{\bf#1},\ #2 (#3)}
\def\RMP #1 #2 #3 {Rev.~Mod.~Phys.~{\bf#1},\ #2 (#3)}
\def\ZP #1 #2 #3 {Z.~Phys.~{\bf#1},\ #2 (#3)}
\def\IJMP #1 #2 #3 {Int.~J.~Mod.~Phys.~{\bf#1},\ #2 (#3)}
\bibliographystyle{unsrt}

{\small
\vspace*{1.00in}
\noindent
Fig.~1: Invariant mass distribution, in 25 GeV wide bins, of the excess
dijet events due to the $Z'$ of the (a)LRM and (b)SSM after QCD background
subtraction at the SSC assuming the same integrated luminosity of 10
$fb^{-1}$. The solid(dash-dotted) curve is the result of performing a best
fit to the excess assuming a Gaussian(Breit-Wigner) shape for these events.

\vspace*{1.00in}
\noindent
Fig.~2: Same as Fig.~1, but for the ER5M $\chi$ assuming an
integrated luminosity of (a) $10 fb^{-1}$ and (b) $100 fb^{-1}$. In the second
case, both Gaussian(solid) and Breit-Wigner(dash-dotted) fits to the peak are
also shown.

\vspace*{1.00in}
\noindent
Fig.~3: Lines of constant $S/{\sqrt {B}}$ in the luminosity-$Z'$ mass
plane. From bottom to top, the lines correspond to $S/{\sqrt {B}}=$2, 3, 4, 5,
6, and 7 for the LRM at the (a) SSC and (b) LHC, for the SSM
at the (c) SSC and (d) LHC, or for the ALRM at the (e) SSC and (f) LHC.

\vspace*{1.00in}
\noindent
Fig.~4: Same as Fig.~3 but for the ER5M $\chi$ at the (a) SSC and (b) LHC, and
for the ER5M $\eta$ at the (c) SSC and (d) LHC.

 }

\end{document}